\documentclass[preprint,12pt]{elsarticle}

\usepackage{graphicx}
\usepackage{amssymb}
\usepackage{amsthm}

\usepackage{amsmath}
\usepackage{caption}
\usepackage{subcaption}
\usepackage{listings}
\usepackage{color}

\journal{Computer Physics Communications}

\begin{document}

\begin{frontmatter}



\title{Event-by-event simulation of a quantum delayed-choice experiment}


\author[RUG]{Hylke C. Donker}
\author[RUG]{Hans De Raedt}
\author[FZJ]{Kristel Michielsen\footnote{Corresponding author: k.michielsen@fz-juelich.de}}

\address[RUG]{Department of Applied Physics,
Zernike Institute for Advanced Materials, University of Groningen, Nijenborgh 4, NL-9747 AG Groningen, The Netherlands}
\address[FZJ]{Institute for Advanced Simulation, J\"ulich Supercomputing Centre, Forschungzentrum J\"ulich, D-52425 J\"ulich, Germany}

\begin{abstract}
The quantum delayed-choice experiment of Tang \emph{et al.} [Nature Photonics \textbf{6} (2012) 600]
is simulated on the level of individual events without
making reference to concepts of quantum theory or without solving a wave equation.
The simulation results are in excellent agreement with the quantum theoretical predictions
of this experiment.
The implication of the work presented in the present paper
is that the experiment of Tang \emph{et al.}
can be explained in terms of cause-and-effect processes in an event-by-event manner.
\end{abstract}



\end{frontmatter}

\section{Introduction}\label{sec:introduction}

\begin{figure}[t]
\begin{center}
\includegraphics[width=0.9\hsize]{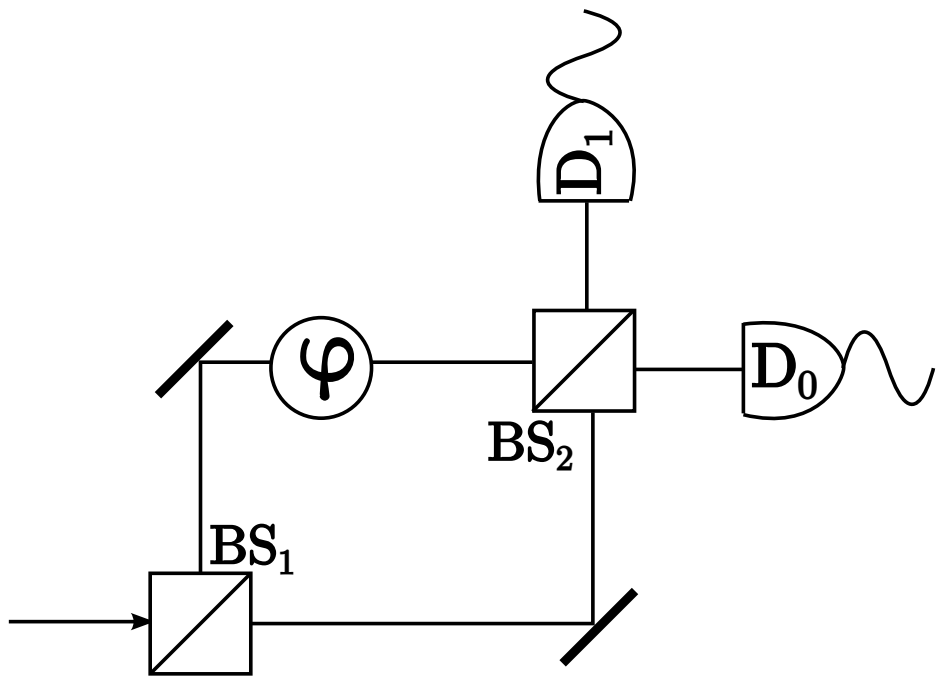}
\end{center}
\caption{Schematic of a
Mach-Zehnder interferometer. BS$_1$ and BS$_2$: beam splitters; $\varphi$: phase shifter; $D_0$ and $D_1$: detectors.}
\label{fig:mzi_setup}
\end{figure}

In a Mach-Zehnder interferometer (MZI) experiment one can choose between measuring the wave-like and the
particle-like properties of photons~\cite{GRAN86,PEDR07}.
The wave-like behavior (interference) is observed using the MZI set-up depicted in Fig.~\ref{fig:mzi_setup}.
The particle-like behavior (no interference) is observed by removing the second beam splitter BS$_2$.
The first and second set-up are henceforth referred to as a closed and an open MZI, respectively.
Hence, the observation of wave-like or particle-like behaviour depends on the choice of considering
a closed or open MZI, respectively, in accordance with the idea of wave-particle duality.
One might therefore ask whether a \emph{normal-choice} and a
\emph{delayed-choice} experiment would yield different observations.
That is, is there a difference in the experimental results if the set-up is already
predetermined to test either the particle or wave nature of a photon (normal-choice) versus a set-up that makes this choice
while the photon has already passed BS$_1$ but not yet BS$_2$ (delayed-choice)?
Experiments have been carried out to measure this difference and there appears to be no difference between these two
situations~\cite{BALD89,HELL87,JACQ07}.

Recently, a new type of delayed-choice experiment has been suggested in which BS$_2$ is a quantum beam splitter assumed to be
in a superposition of being present and absent~\cite{IONI11}.
This so-called \emph{quantum delayed-choice} experiment
has been realized experimentally using NMR interferometry on ensembles of molecules~\cite{ROY12, AUCC12}
and using single-photon quantum optics techniques~\cite{PERU12, TANG12, KAIS12}.
These experiments demonstrate that in one single experiment, particle- or wave-like behavior can be tuned continuously,
which is interpreted as an indication that the complementarity principle needs refinement~\cite{PERU12, TANG12, KAIS12}.

From the viewpoint of quantum theory, the central issue is how it can be that
experiments such as these delayed-choice experiments yield definite answers~\cite{LEGG87}.
As the concept of an event is not a part of quantum theory proper,
quantum theory simply cannot address the question ``why there are events?''~\cite{ENGL13}.
One can get around this conundrum by constructing a description entirely in terms of events,
ultimately related to human experience, and the cause-and-effect relations among them.
Such an event-based description obviously yields definite answers and if
it reproduces the statistical results of experiments,
it also provides a description of the experiments on a level of detail that is not covered by quantum theory.

Essentially, the event-based approach is based on the fact that all
that what can be said about nature is constrained by the data a measurement apparatus can,
at least in principle, produce. As Wheeler put it:
``[...] every particle, every field of force, even the space-time continuum itself--derives its function,
its meaning, its very existence entirely [...] from the apparatus-elicited answers to yes-or-no questions [...].''~\cite{WHEE89}.
The event-based approach is to be viewed in this light;
An event-based simulation does not necessarily mimic what actually happens in nature:
it only produces sets of data (e.g. detector clicks) that can be compared to experiments in the laboratory
through a chronological, causally-connected sequence of events.
From this it directly follows that such an event-based approach has no bearing on
the interpretation, applicability, validity or possible extensions of quantum theory.

For many interference and entanglement phenomena observed in quantum-optics and single-neutron experiments,
such an event-based description has already been constructed~\cite{RAED05d,RAED05b,MICH11a,RAED12a,RAED12b}.
The event-based simulation approach reproduces the statistical distributions of quantum theory
by modeling physical phenomena as a chronological sequence of events, by neither solving a wave equation
nor by sampling a distribution as in a Monte-Carlo simulation.
Hereby events can be actions of an experimenter, particle emissions by a source, signal generations by a detector,
interactions of a particle with a material and so on~\cite{MICH11a,RAED12a,RAED12b}.

In the context of the work presented in this paper
we mention that the event-by-event simulations have successfully been used to
reproduce the results of the single-photon MZI experiment of Grangier {\it et al.}~\cite{GRAN86} (see Refs.~\cite{RAED05d,MICH11a}),
the single-photon Wheeler delayed choice experiment
by Jacques {\it et al.}~\cite{JACQ07,JACQ08} (see Refs.~\cite{MICH11a,ZHAO08b})
and the proposal for a quantum delayed-choice experiment~\cite{IONI11} in terms of quantum gates~\cite{RAED12e},
thereby employing the event-based method to simulate a universal quantum computer~\cite{MICH05}.

In this paper we demonstrate that results of the single-photon quantum delayed-choice experiment~\cite{TANG12},
a so-called quantum-controlled experiment because conceptually it involves controlling the presence/absence of a beam splitter
by a qubit,
can be reproduced by an event-based model that is a one-to-one copy of the actual experiment.
The event-based simulation is Einstein-local and causal and does not rely on concepts of quantum theory.
Therefore, in contrast to the general belief~\cite{ANAN13}, both the quantum delayed-choice experiment~\cite{IONI11} and
Wheeler's delayed choice experiment can be explained entirely in terms of particle-like objects travelling one-by-one through the
experimental set-up and generating clicks of a detector, thereby providing a mystery-free explanation
of the experimental results.

\section{Quantum theoretical description}
\begin{figure}
\begin{center}
\includegraphics[width=0.9\hsize]{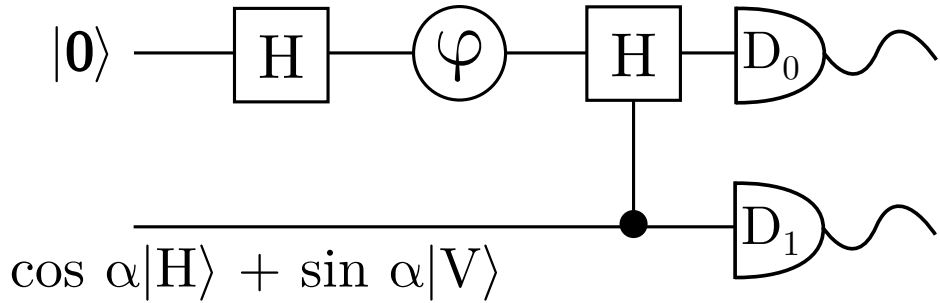}
\end{center}
\caption{Quantum network of a quantum
delayed-choice experiment. The first Hadamard (H) gate, corresponding to BS$_1$ in Fig. \ref{fig:mzi_setup}, is followed by a phase
shifter $\varphi$ and a second Hadamard gate, corresponding to BS$_2$ in Fig. \ref{fig:mzi_setup}.
BS$_2$ can be set in a superposition of being present and absent by controlling the state of an ancilla.
The photon and ancilla are detected by detectors $D_0$ and $D_1$, respectively, after the control operation on the second Hadamard gate. The
photon and the ancilla are prepared in the state $|0\rangle$ and $\cos\alpha|H\rangle + \sin \alpha|V\rangle$, respectively.}
\label{fig:quantum_network}
\end{figure}

Conceptually, the quantum delayed-choice experiment~\cite{TANG12} is conveniently represented by a quantum-gate network,
see Fig.\ref{fig:quantum_network}.
The first Hadamard operation, equivalent to the operation of beam splitter BS$_1$, transforms the initial state
$|0\rangle$ into the superposition $(|0\rangle+|1\rangle )/\sqrt{2}$, where $|0\rangle$ and $|1\rangle$ represent the optical paths
(spatial modes) of the photon in the MZI.
The phase shifter $\varphi$ changes the relative phase between the optical paths.
This results in the spatial state $|{\rm space}\rangle =(|0\rangle+e^{i\varphi}|1\rangle )/\sqrt{2}$ of the photon.
In the quantum delayed-choice experiment beam splitter BS$_2$ is controlled by an ancilla
and can be in a superposition of being present and absent.
In the experimental realization~\cite{TANG12} the polarization state $|{\rm pol}\rangle$
of the photon is taken to be the ancilla.
If the photon is horizontally polarized ($|{\rm pol}\rangle=|H\rangle $),
then the photon can pass BS$_2$ (closed MZI) and if it is vertically
polarized ($|{\rm pol}\rangle=|V\rangle $) then it cannot pass BS$_2$ (open MZI).
Hence, BS$_2$ is a polarization controlled beam splitter.

If the ancilla is prepared in the state $|{\rm pol}\rangle = \cos\alpha|H\rangle + \sin \alpha|V\rangle $, where $\alpha$
denotes the polarization angle of the photon, then the total state of the photon before arriving at BS$_2$ reads
$|\psi\rangle =|{\rm space}\rangle |{\rm pol}\rangle = (|0\rangle+e^{i\varphi}|1\rangle )(\cos\alpha|H\rangle + \sin \alpha|V\rangle )/\sqrt{2}$.
After the operation of the second Hadamard gate (BS$_2$) this state becomes
$|\psi\rangle =\sin\alpha |{\rm particle}\rangle |V\rangle +\cos\alpha |{\rm wave}\rangle |H\rangle$, where
$|{\rm particle }\rangle = (|0\rangle+e^{i\varphi}|1\rangle )/\sqrt{2}$ and
$|{\rm wave}\rangle = e^{i\varphi /2}\left(e^{i\delta_0}\cos\frac{\varphi}{2}|0\rangle -ie^{i\delta_1}\sin\frac{\varphi}{2}|1\rangle\right)$ describes
wave-like behavior. The extra phase shifts $\delta_0$ and $\delta_1$ originate from the specific experimental set-up~\cite{TANG12}.

There are now two ways to proceed.
First, considering the polarization states $|H\rangle$ and $|V\rangle$ in $|\psi\rangle$ as a label for the particle and wave properties,
a classical mixture of these properties is described by the mixed state
$\rho = \sin^2\alpha |{\rm particle}\rangle\langle {\rm particle}|+ \cos^2\alpha |{\rm wave}\rangle\langle {\rm wave}|$.
This corresponds to Wheeler's delayed choice experiment~\cite{JACQ07,JACQ08}.
In this case the normalized intensities at detectors $D_0$ and $D_1$ are given by
\begin{equation}
I_0=(1+\cos^2\alpha\cos\varphi )/2,
\label{I0}
\end{equation}
and
\begin{equation}
I_1=(1-\cos^2\alpha\cos\varphi )/2,
\label{I1}
\end{equation}
respectively.
Second, by measuring the ancilla in the $|+\rangle=(|H\rangle + |V\rangle )/\sqrt{2}$ basis the photon state becomes
$|\psi \rangle =\sin\alpha |{\rm particle}\rangle +\cos\alpha |{\rm wave}\rangle$.
This measurement can be performed by placing 45$^{\circ}$ polarizers in between BS$_2$ and the detectors $D_0$ and $D_1$.
In this quantum delayed-choice experiment the non-normalized intensities at detectors $D_0$ and $D_1$ are given by~\cite{TANG12}
\begin{equation}
I_0=\frac{1}{4}+\frac{1}{4}\cos^2\alpha\cos\varphi+\frac{\sqrt{2}}{4}\sin2\alpha\cos\frac{\varphi}{2}\cos\left(\frac{\varphi}{2}+\delta_0\right),
\label{I0Q}
\end{equation}
and
\begin{equation}
I_1=\frac{1}{4}-\frac{1}{4}\cos^2\alpha\cos\varphi-\frac{\sqrt{2}}{4}\sin2\alpha\sin\frac{\varphi}{2}\sin\left(\frac{\varphi}{2}-\delta_1\right),
\label{I1Q}
\end{equation}
respectively.
The normalized intensities then read $I_0/(I_0+I_1)$ and $I_1/(I_0+I_1)$, respectively.

\begin{figure*}[t]
\begin{center}
\includegraphics[width=0.95\hsize]{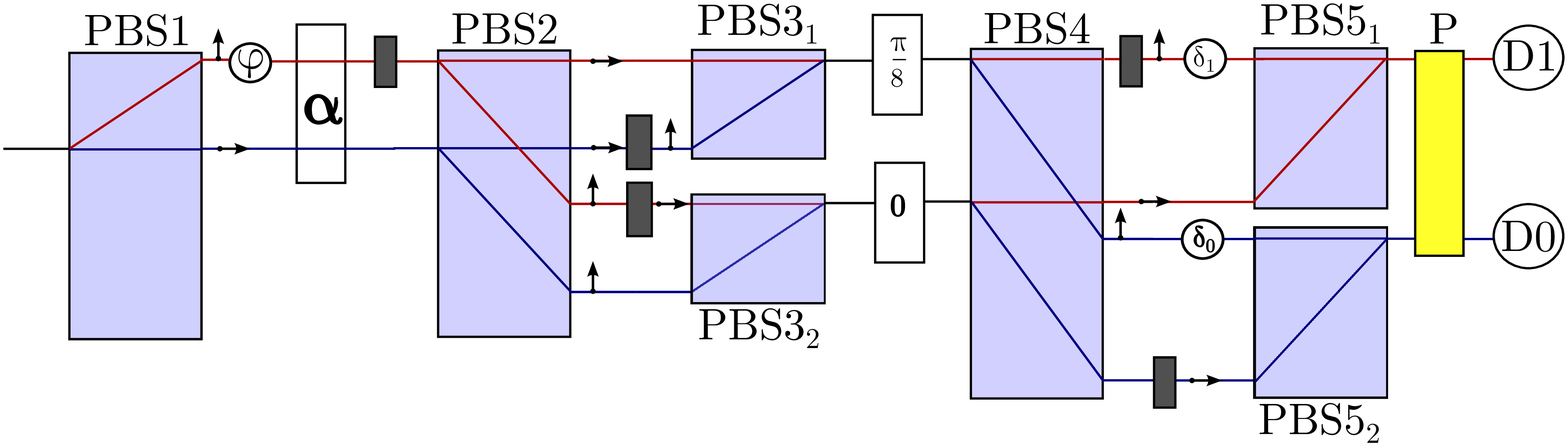}
\end{center}
\caption{Schematic of the quantum delayed choice experiment by Tang {\sl et al.}~\cite{TANG12}
and of the event-based simulation model.
PBS$X_{(y)}$ with $X=1,\dots,5$ and $y=1,2$: polarizing beam splitters; $\varphi$, $\delta_0$ and $\delta_1$: phase shifters; grey
rectangles: half wave plates, white rectangles: half wave plates at angles $\alpha$, 0 and $\pi /8$; P: polarizer;
$D_0$ and $D_1$: detectors. The arrows indicate the polarization direction of the photons.} \label{fig:set_up}
\end{figure*}

\section{Experimental realization}\label{sec:exp}

The diagram of the experimental set-up used by Tang {\sl et al.}~\cite{TANG12} is shown in Fig.~\ref{fig:set_up}.
Except for the photon source it depicts all the optical components that are being used.
The source, an InAs/GaAs self-assembled quantum dot~\cite{TANG12}, emits polarized single photons.
The polarization state $|H\rangle$ ($|V\rangle$) of the photon
is used as the ancilla and plays the role of the control qubit represented in Fig.~\ref{fig:quantum_network} by the bottom line.
The path of the photon corresponds to the target qubit, represented by the top line in Fig.~\ref{fig:quantum_network}.

The correspondence between the components of the quantum network Fig.~\ref{fig:quantum_network}
and the experimental realization depicted in Fig.~\ref{fig:set_up} is as follows
(throughout we adopt the quantum theoretical, wave mechanical picture).
The first polarizing beam splitter PBS1 and the phase shifter $\varphi$
perform the first Hadamard and phase shift operation.
The half wave plate (HWP) with angle $\alpha$ creates
the superposition of both an open and a closed MZI,
that is it prepares the ancilla in the state $|{\rm pol}\rangle = \cos\alpha|H\rangle + \sin \alpha|V\rangle $.
The HWP behind the HWP with angle $\alpha$ (and all other HWPs represented by the same graphic symbol)
serves as a compensator: they have no equivalent in the quantum network Fig.~\ref{fig:quantum_network} but they are essential
to the physical realization of the experiment.
Depending on the polarization state of the photon, PBS2 directs the photons (i.e. split the probability wave)
in an upper pair and lower pair of paths.
These two paths correspond to the two possible states of the target qubit in Fig.~\ref{fig:quantum_network}.
The upper(bottom) pair of paths realizes the closed(open) MZI.
Note that two lines within each pair correspond to the polarization state of the photon,
i.e. they represent the two possible states of the ancilla in the quantum network.
The controlled Hadamard gate in Fig.~\ref{fig:quantum_network} is realized
by the components PBS3$_1$, PBS3$_2$, the HWP's at angles 0 and $\pi /8$,
PBS4, PBS5$_1$ and PBS5$_2$.
The paths of the open and closed MZI merge at PBS5.
Up to this point in the diagram in Fig.~\ref{fig:set_up}, the successive operations of the optical components on the
photons implements the action of the quantum gates in Fig.~\ref{fig:quantum_network} on a 2-qubit system.

In the work of Tang {\sl et al.}~\cite{TANG12}, two cases are considered.
In the first case, the polarizer P is in place (see Fig.~\ref{fig:set_up})
and points in the $\left[1,1\right]^T/\sqrt{2}$ direction.
A photon leaving the polarizer is counted by either detector $D_0$ or $D_1$.
In the second case, the polarizer P is absent and
photons passing through PBS5$_1$ (PBS5$_2$) are counted at detector $D_1$ ($D_0$).
In our simulation work, we assume that the single-photon detectors have 100\%
detection efficiency, that is we count every photon that is emitted by the source.

\section{Event-based simulation}\label{sec:computational_model}

The simulation model is a one-to-one copy of the quantum delayed-choice experiment~\cite{TANG12}.
A single photon is represented by a messenger that travels through the network~\cite{MICH11a,RAED12e},
the diagram of which is a copy of the experimental set-up and hence is the same as Fig.~\ref{fig:set_up}.
The creation of a photon by the source corresponds to an event that creates a messenger.
The message carried by the messenger consists of a two-dimensional, complex-valued unit vector~\cite{MICH11a,RAED12e},
representing the two electric field components which are orthogonal to the direction of propagation.
It may help to think of the messenger as carrying a normalized Jones vector~\cite{PEDR07}.

In the event-based approach, optical components are represented by processing units
which accept a messenger, interpret and transform its messages and send out
a messenger with its modified message~\cite{ZHAO08b,MICH11a,RAED12e}.
The frequency distributions of detection events are generated event-by-event,
meaning that at any time only a single messenger travels through the network of
components whilst at any given time it has a definite position.
Hence the single clicks that are produced by the detectors (which in our simulation are assumed
to count every event that impinges on the detectors)
build up the interference pattern one-by-one, exactly as in the laboratory experiment.

In the present work, the optical components are represented by exactly the same processing units
as the ones used in earlier work~\cite{ZHAO08b,MICH11a,RAED12e}.
For completeness, we briefly describe the action of each processing unit that appears
in Fig.~\ref{fig:set_up}.
The Python code listed in the Appendix provides the complete specification the algorithm that
we have used to produce the data reported in the present paper.

\textit{Polarizing beam splitter (PBS):} This optical component is represented by a two-input -- two-output processor.
A messenger arriving at one of its input ports causes the internal state (represented by ten floating-point numbers)
of the processor to be updated.
The update rule consists of storing the message into memory local to the processor
and changing a two-dimensional vector, also local to the processor, which serves to estimate the
relative number of messengers arriving at the two different input ports.

Pictorially, the update rule mimics the change of the polarization of the dielectric
due to the interaction of the latter with electric field~\cite{RAED12b}.
The ten numbers contained in the internal state are used to build a unit vector which is then multiplied
by the unitary transformation which is the same as the one used in the classical (and quantum) wave mechanical description
of a polarizing beam splitter~\cite{GARR08,BACH08}.
The resulting unit vector is then used to (i) decide at which output port the messenger
will leave the unit and (ii) to construct the new message which the messenger will carry.
The technical details of examples of the implementation of the algorithm that we have just described
are given in Refs.~~\cite{ZHAO08b,MICH11a,RAED12e}.
It is important to note that the processing unit does not store information from all
the messengers that it processes: it only has a very limited amount of memory, namely
ten floating point numbers only.

Since both the memory of the processing unit and a (pseudo) random number
is taken into account to generate an output message,
there is no one-to-one relation between the input and the output message.
Hence, this processing unit is indeed different from,
and therefore there is also no correspondence with, the unitary time evolution of quantum theory.
Unlike the suggestion made in Fig.~\ref{fig:set_up}, in the event-based simulation
both PBS2 and PBS4 are represented by two identical but logically independent
processing units.

\textit{Phase shifters:} In the event-based approach, phase shifters change the time of flight
of the messenger, corresponding to a rotation of the unit vector encoding the message
by e.g. $\varphi$, $\delta_0$ or $\delta_1$.

\textit{Polarizer:} The polarizer is modeled as a PBS which is rotated about 45 degrees,
corresponding to the change of the basis from $(|H\rangle,|V\rangle)$ to
$((|H\rangle+|V\rangle)/\sqrt{2},(|H\rangle-|V\rangle)/\sqrt{2})$,
and by destroying all messengers which leave the PBS via one of the output ports and keeping
all others.

\textit{Half wave plate (HWP):} a HWP is modeled as a passive device that rotates the message
according to the unitary matrix which is characterized by $\theta$, the angle between the
optical axis and the laboratory frame~\cite{MICH11a, PEDR07}.

Looking at Fig.~\ref{fig:set_up}, the solid rectangles represent HWPs
with $\theta = \pi/4$ whilst the other HWPs have angles $\theta$ as indicated.

\textit{Detectors:} In our simulations, the detectors are considered to be ideal.
This means that whenever a messenger reaches the detector, the detector count is incremented by one.

\section{Simulation procedure}
As in the experiment~\cite{TANG12}, we perform two sets of simulations:
One with and one without the polarizer in front of the detectors (see Fig.~\ref{fig:set_up}).
For the angle of polarization we take $\alpha = l\pi/8$ for $l=0,\dots,7$,
the values for which Ref.~\cite{TANG12} reports experimental data.
As Ref.~\cite{TANG12} does not mention the values of the phase shifts
$\delta_0$ and $\delta_1$, we adjust these parameters such that the quantum theoretical
and event-based simulation results look very similar to the experimental data.
The results presented here have been obtained by taking $\delta_0=\pi/8$ and $\delta_1=-7\pi/40$.
Note that quantum theory predicts that the choice of these two parameters only affects the results if the polarizer P is present
(compare Eqs.~(\ref{I0},\ref{I1}) and Eqs.~(\ref{I0Q},\ref{I1Q})).

For each set of values of the phase shift $\varphi$ and rotation angle $\alpha$,
the internal states of the processing units are initialized by means of different pseudo-random numbers.
The data collection consists of sending, one-by-one, $N=10000$ messengers through the network.
Only after a messenger has been processed by a detector unit or has been destroyed by the polarizer P,
a new messenger is being created.
At creation, the messenger carries as message the unit vector $\mathbf{v} =\left[1, 1\right]^T/\sqrt{2}$
corresponding to a linear polarization of 45 degrees.

Every messenger that arrives at a detector increases the corresponding event count by one.
The relative detection count is obtained by dividing the event count by the
total number of events registered by both detectors.

\section{Simulation results}

\begin{figure}[h]
\begin{center}
\begin{subfigure}[b]{0.4\hsize}
\includegraphics[width=1.0\hsize]{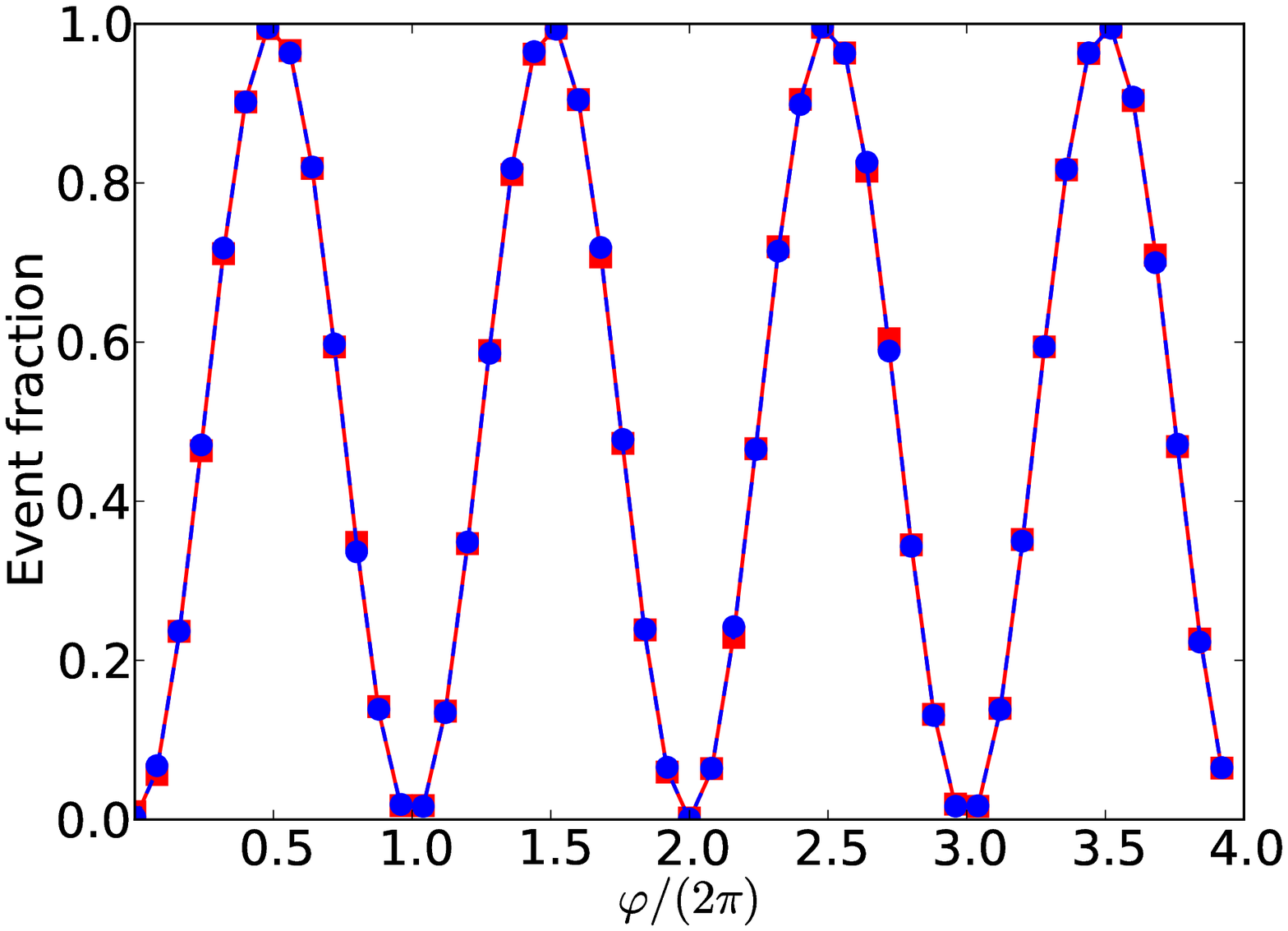}
\caption{$\alpha = 0$}
\label{fig:quantum_wdc0}
\end{subfigure}
\begin{subfigure}[b]{0.4\hsize}
\includegraphics[width=\hsize]{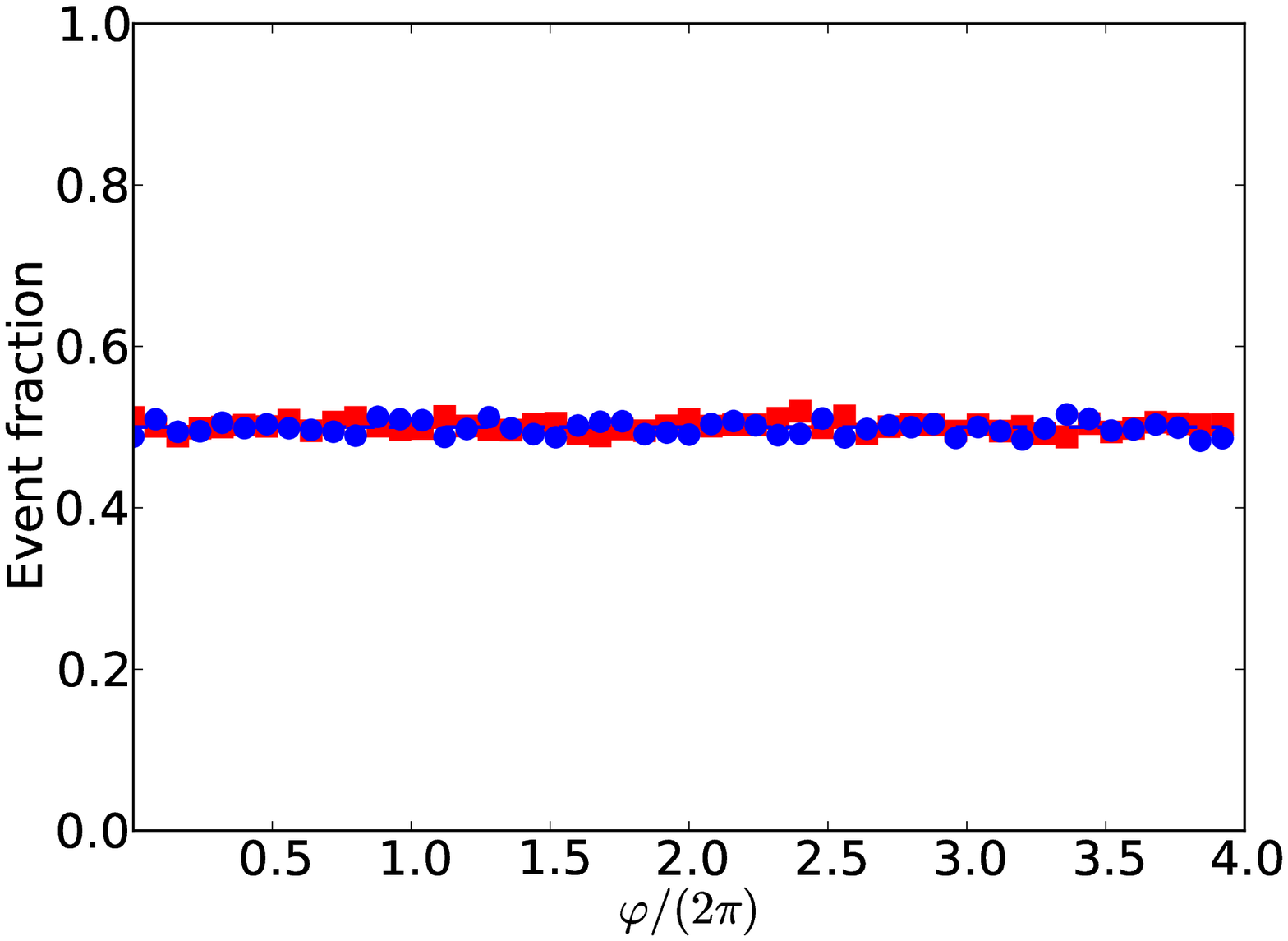}
\caption{$\alpha = \pi/2$}
\label{fig:quantum_wdc4}
\end{subfigure}
\begin{subfigure}[b]{0.4\hsize}
\includegraphics[width=\hsize]{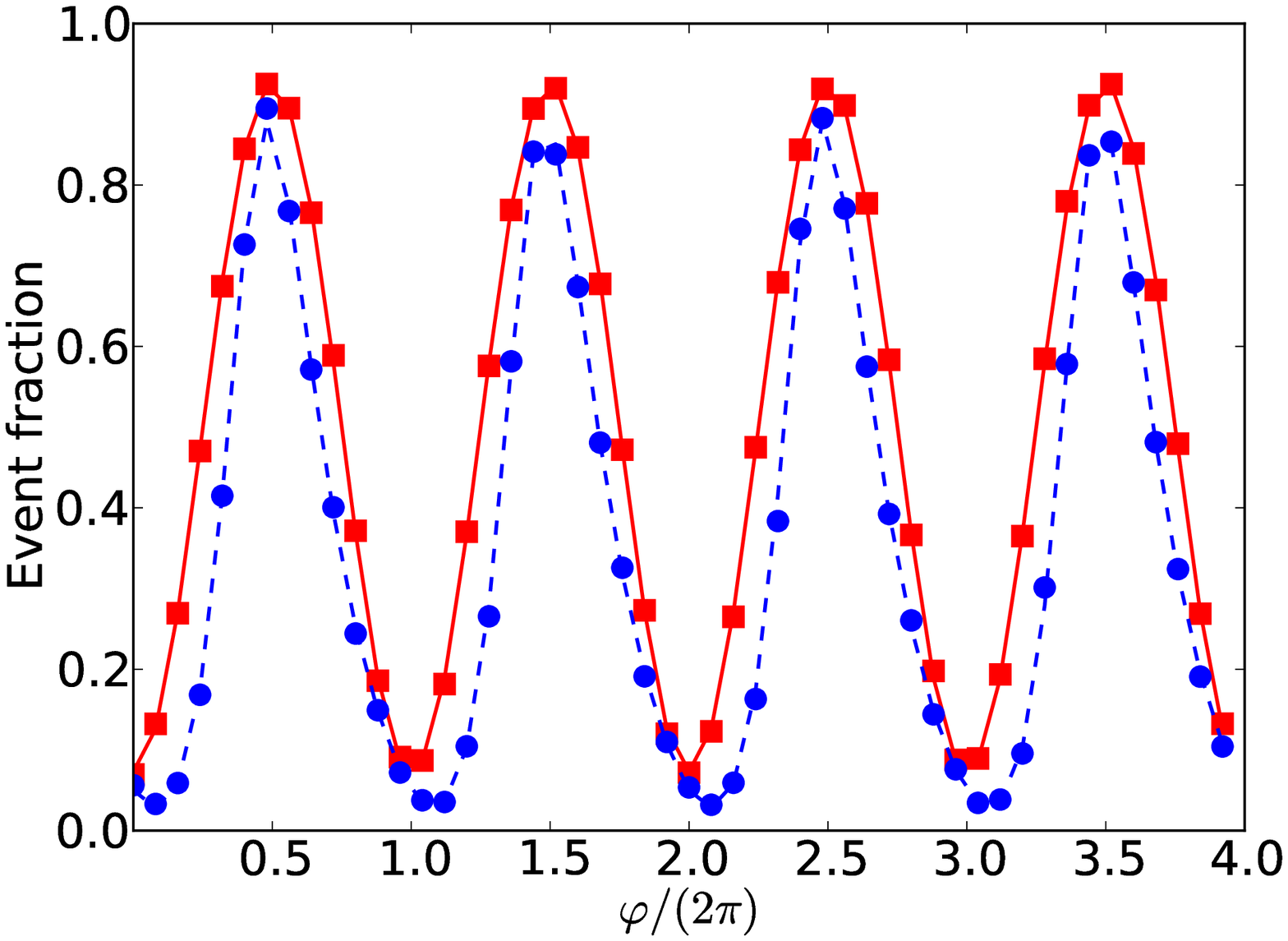}
\caption{$\alpha = \pi/8$}
\label{fig:quantum_wdc1}
\end{subfigure}
\begin{subfigure}[b]{0.4\hsize}
\includegraphics[width=\hsize]{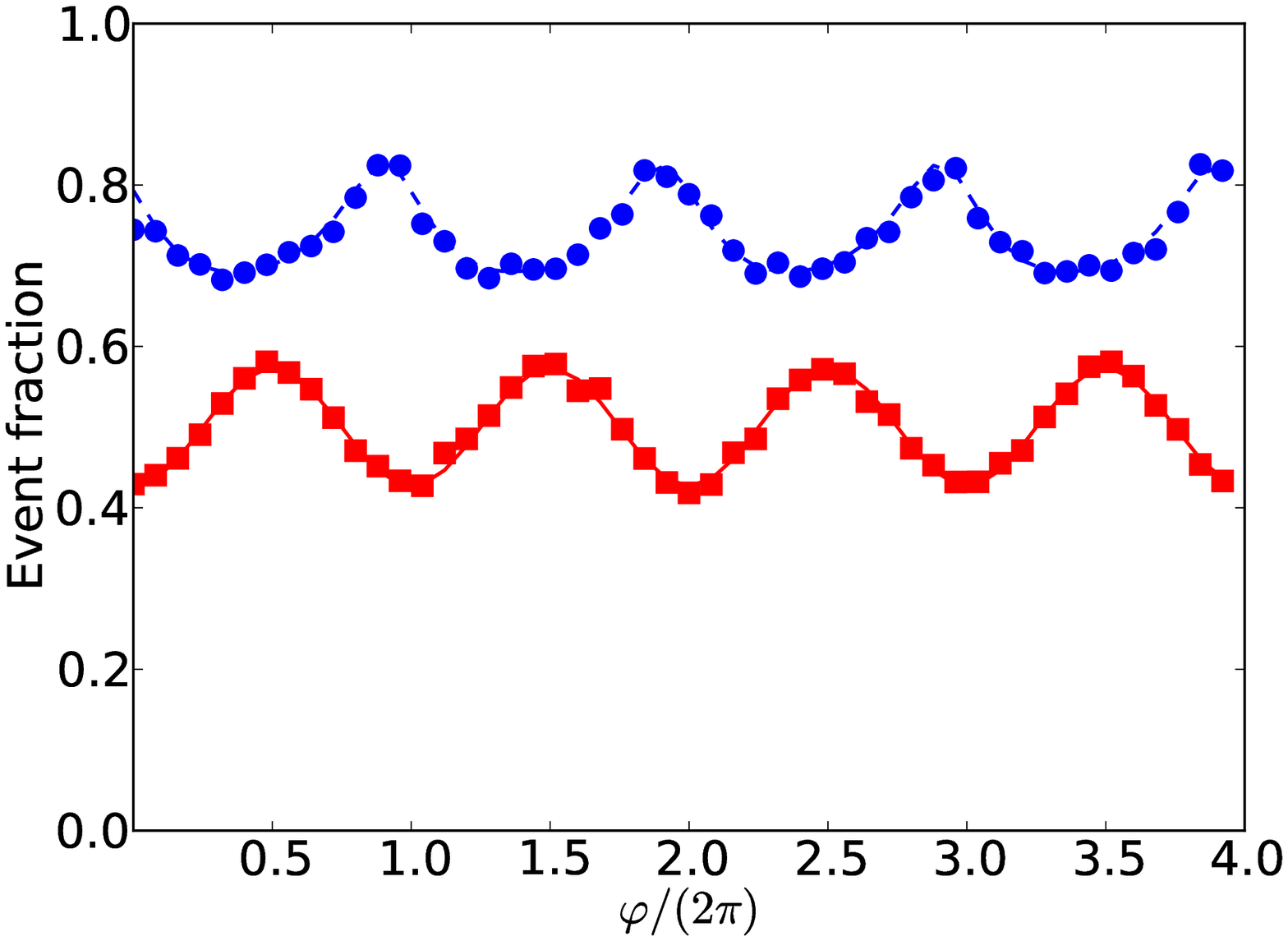}
\caption{$\alpha = 5\pi/8$}
\label{fig:quantum_wdc5}
\end{subfigure}
\begin{subfigure}[b]{0.4\hsize}
\includegraphics[width=\hsize]{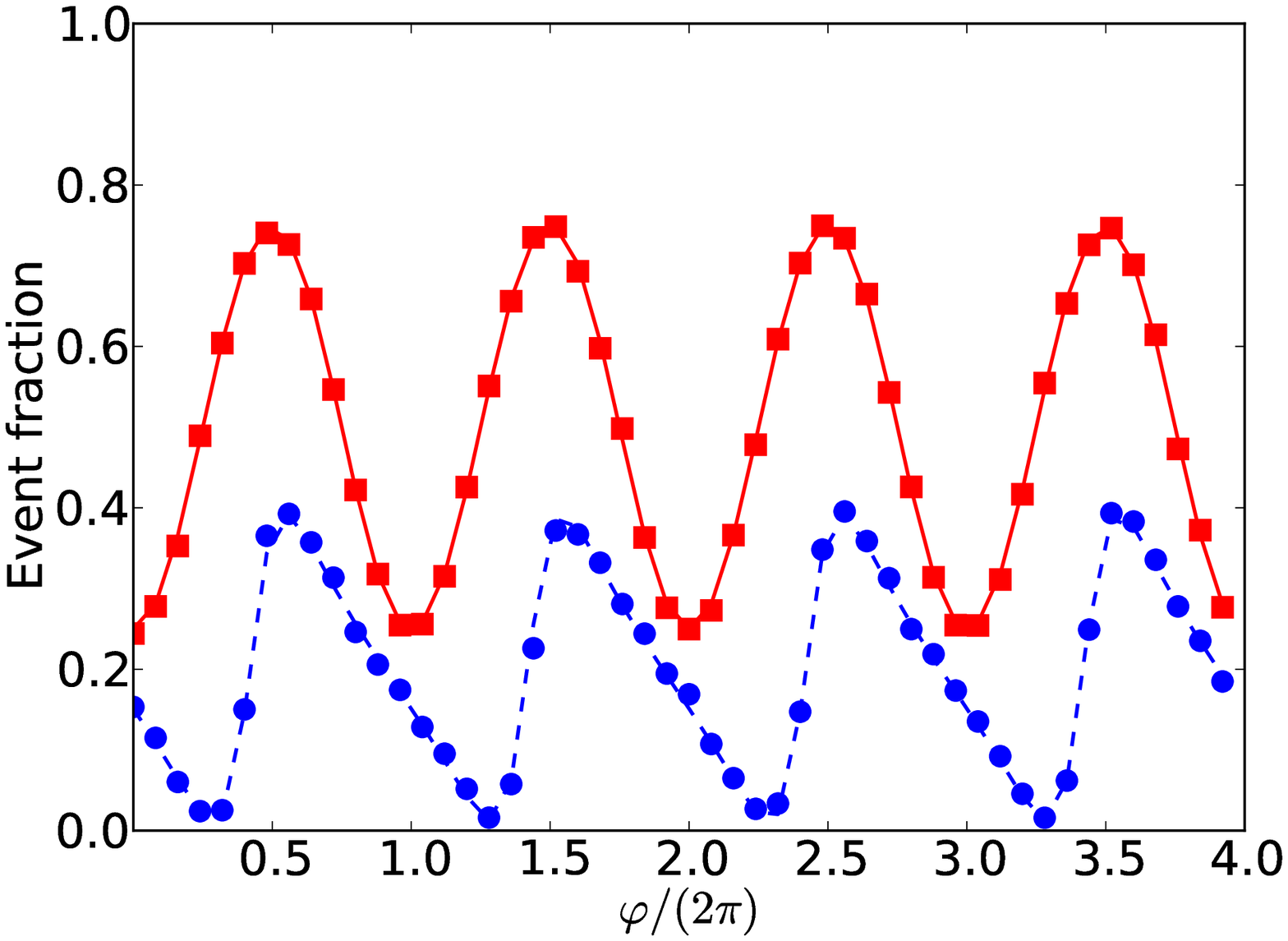}
\caption{$\alpha = \pi/4$}
\label{fig:quantum_wdc2}
\end{subfigure}
\begin{subfigure}[b]{0.4\hsize}
\includegraphics[width=\hsize]{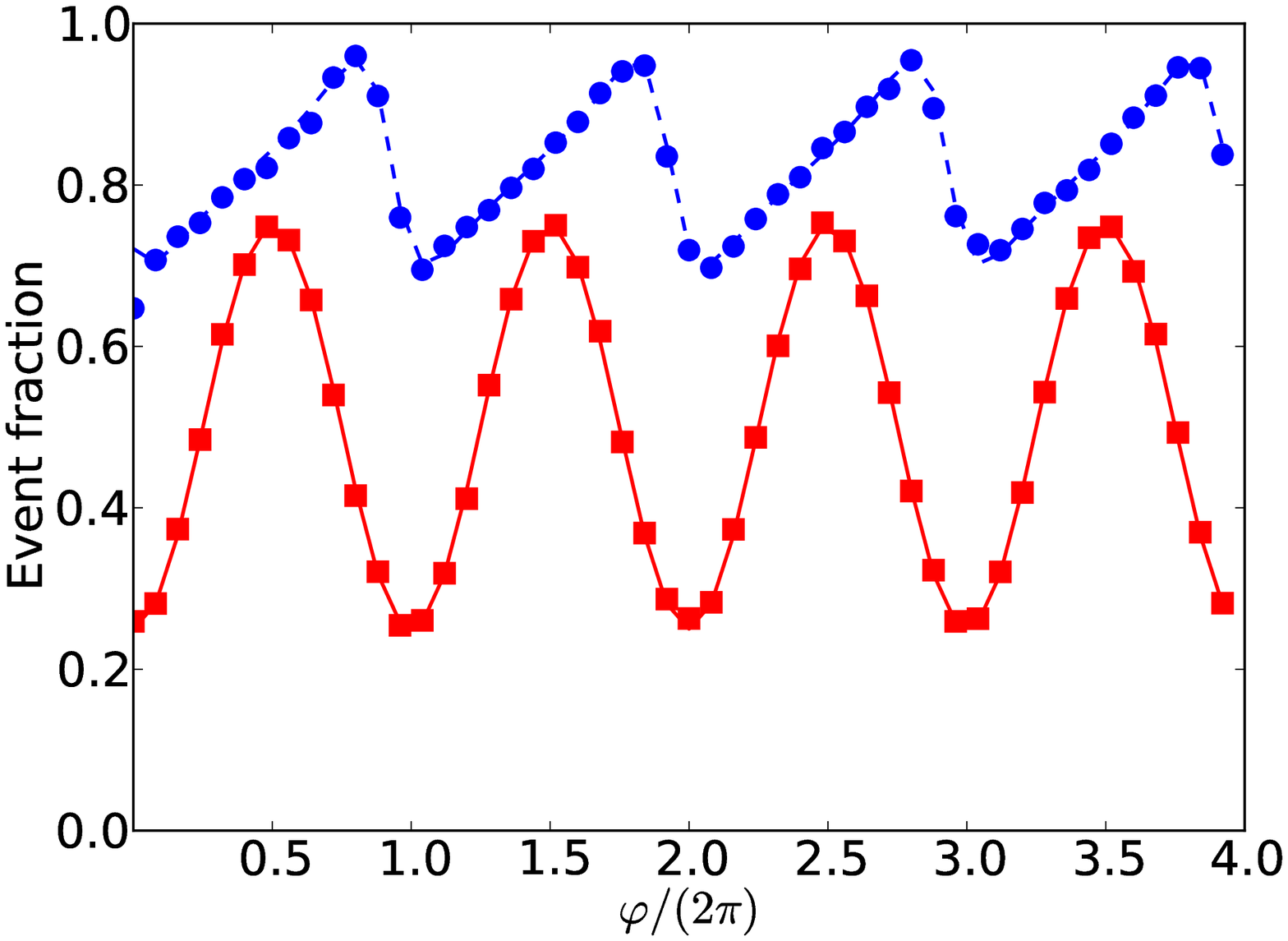}
\caption{$\alpha = 3\pi/4$}
\label{fig:quantum_wdc6}
\end{subfigure}
\begin{subfigure}[b]{0.4\hsize}
\includegraphics[width=\hsize]{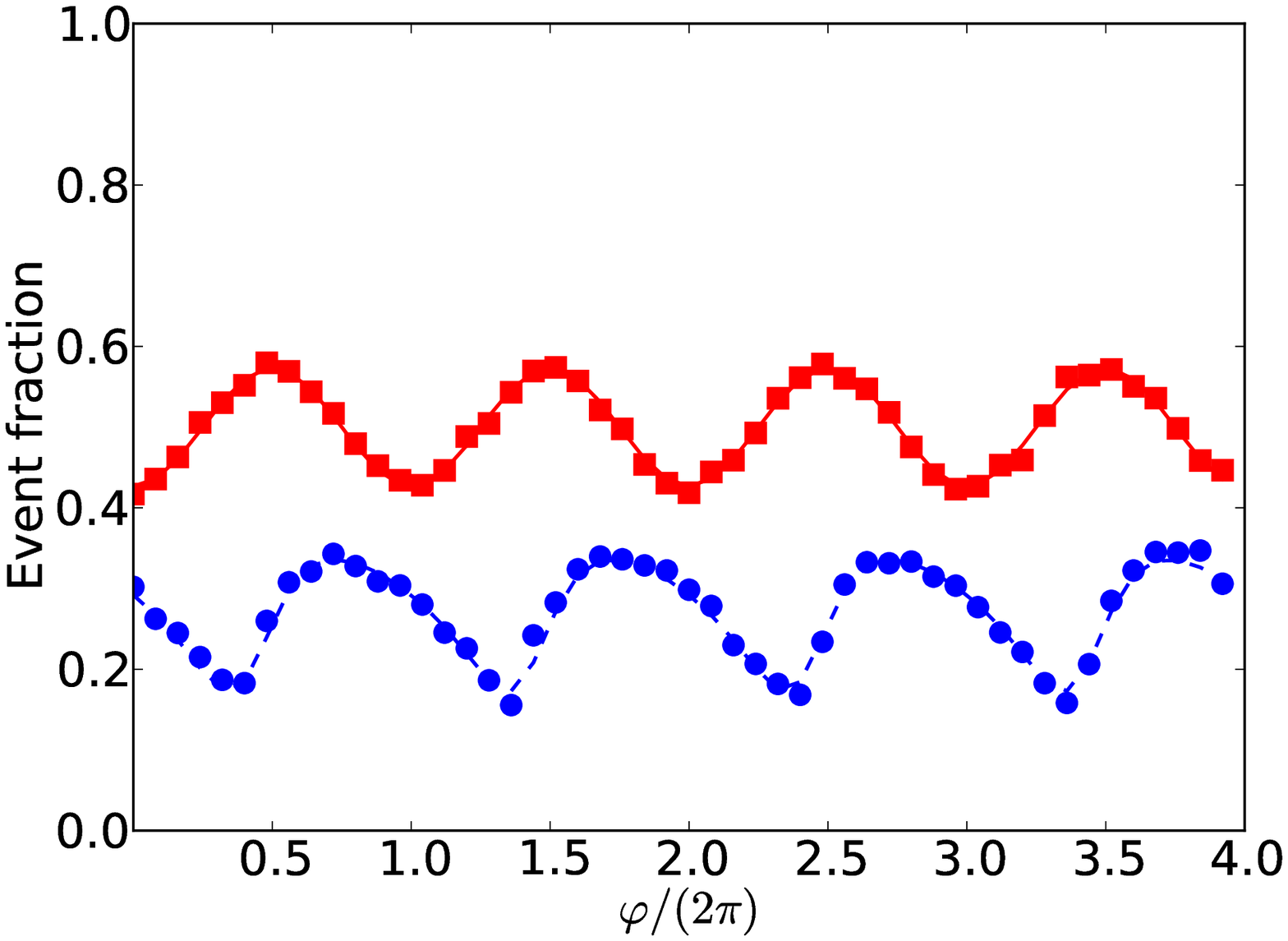}
\caption{$\alpha = 3\pi/8$}
\label{fig:quantum_wdc3}
\end{subfigure}
\begin{subfigure}[b]{0.4\hsize}
\includegraphics[width=\hsize]{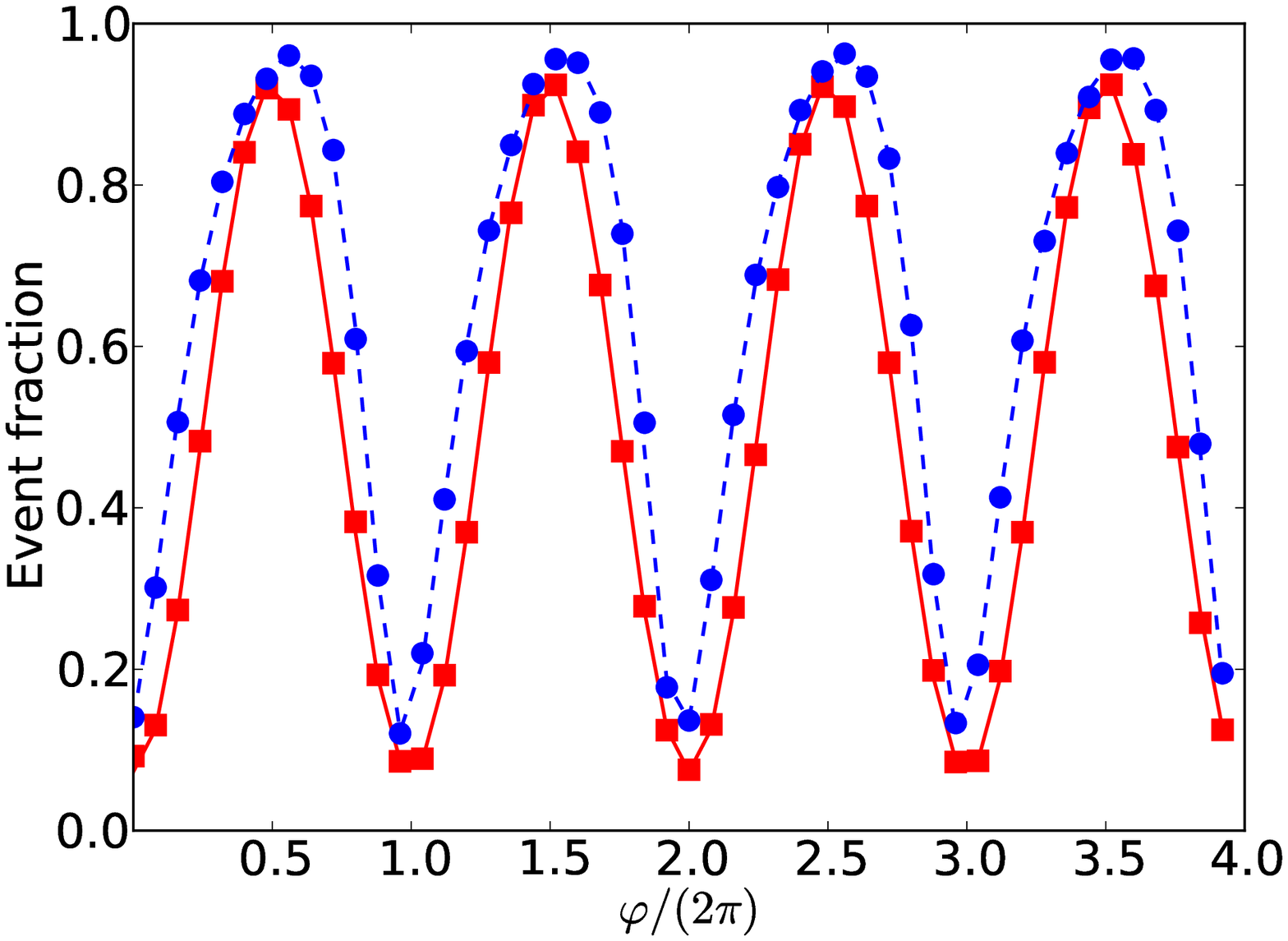}
\caption{$\alpha = 7\pi/8$}
\label{fig:quantum_wdc7}
\end{subfigure}
\caption{Fraction of events detected at detector $D_1$ (markers) as a function of the phase difference $\varphi$ for various polarization
angles $\alpha$ in case of a Wheeler's delayed choice type of experiment (red squares) and a quantum delayed choice type of experiment
(blue circles). For each set of $(\varphi,\alpha)$, the number of input events (photons) $N=10000$;
phase shift $\delta_0 = \pi/8 $; phase shift $\delta_1 = -7\pi/40$.
The lines show the corresponding quantum theoretical results for the normalized intensity $I_1$ derived from
Eq.~(\ref{I1}) (red solid line) and Eq.~(\ref{I1Q}) (blue dashed line), respectively.}
\label{fig:wdc_results}
\end{center}
\end{figure}

The simulation procedure outlined above is repeated for 50 different values of the phase shift $\varphi$,
yielding the plots presented in Fig.~~\ref{fig:wdc_results}.
The fraction of events arriving at $D_1$ is plotted as a function of the phase difference $\varphi$.
The (red) squares represent the event-based simulation data for the set-up without the polarizer P and the (red) solid lines
are the corresponding quantum theoretical predictions (see Eq.~(\ref{I1})).
The (blue) circles represent the event-based simulation results
for the set-up with the polarizer P present and the (blue) dashed lines are the predictions according to quantum theory (Eq.~(\ref{I1Q})).
Each subpanel shows the results for a different value of $\alpha$.
In all cases there is excellent agreement between the simulation data and
the quantum theoretical description.
Therefore, we have established that the ``morphing'' of
the particle and wave nature of the photon in a quantum delayed-choice experiment~\cite{TANG12}
can be explained without recourse to concepts of quantum theory.

\clearpage

\section{Discussion}

As mentioned in the introduction, the event-based approach has successfully been applied
to a large variety of single-photon and single-neutron experiments that involve
interference and entanglement.
In the present paper, we have shown that, without any modification, the same simulation approach can also mimic,
event-by-event, a quantum delayed choice experiment.
As none of these demonstrations rely on concepts of quantum theory
and as it is unlikely that the success of all these demonstrations is accidental,
one may wonder what it is that makes a system genuine ``quantum''.
Indeed, if a physicist living before the era of quantum physics was asked to calculate
the intensities for the set-up depicted in Fig.~\ref{fig:set_up} he would have had no problems
finding the correct expressions using classical electrodynamics.
What makes this experiment quantum (as opposed to classical) is the fact that single indivisible entities
called photons traverse the apparatus, one single quantum at a time.
Yet, paradoxically this is precisely what quantum theory cannot describe since it
can only describe the statistical distribution, not the single events.
The event-by-event approach fills in this gap by combining classical electrodynamics with the notion of particles
(which is not part of Maxwell's theory).
The key to produce interference patterns without having to rely on a wave equation
is the introduction of an adaptive processing element which simulates the interaction
between single particles and the optical components,
akin to the classical electrodynamics description of the response of
the polarization of the material to the applied electric field.
There are several different ways to implement
such adaptive processing elements and although the details of their
dynamics is different, in the stationary state they yield the same frequency
distributions for observing the events~\cite{RAED05d,MICH11a}.
In order to determine which of these implementations is closer to what
actually happens in experiments (on the level of individual events),
it is necessary to perform new experiments which specifically address this question.

To further address the question what makes a certain system genuine ``quantum''
it is of interest to recall Bohr's point of view
that ``There is no quantum world. There is only an abstract physical description''
(reported by~\cite{PETE63}, for a discussion see~\cite{PLOT10a})
and that ``The physical content of quantum mechanics is exhausted
by its power to formulate statistical laws''~\cite{BOHR99}.
Or, to say it differently
``Quantum theory describes our knowledge of the atomic phenomena rather than
the atomic phenomena themselves''~\cite{LAUR97}.
In other words, quantum theory captures,
and does so extremely well, the inferences that we, humans,
make on the basis of experimental data~\cite{RAED14b}.
However it does not describe cause-and-effect processes.
Quantum theory predicts the probabilities that events occur,
but it cannot answer the question ``Why are there events?''~\cite{ENGL13}.
The logical contradictions, often referred to as mysteries~\cite{ANAN13},
created by the (quantum) delayed choice experiment
are a direct consequence of trying to explain the existence of events
within the realm of quantum theory proper.
The point of view that we take in developing event-based descriptions is that
on a basic level, it is our perceptual and cognitive system that defines, registers and processes events.
Therefore, events themselves and the rules that create new events are taken to be the key elements
in the construction of the event-based models.

In conclusion, the implication of the work presented in the present paper
is that the results of the beautiful quantum-delayed choice experiment of Tang \emph{et al.}~\cite{TANG12}
can be explained entirely in terms of cause-and-effect, discrete-event processes,
without making reference to quantum or wave theory and invoking concepts
and ideas that defy common sense~\cite{ANAN13}.
In particular, it is not necessary to introduce the idea of morphing between
the particle- and wave-like behavior of photons~\cite{IONI11}, a particle-only description
suffices to account for the experimental observations.

\section{Acknowledgement}
K.M. would like to thank Pablo Manuel Cencillo Abad for exploratory research work.

\section*{Appendix: Python program}
The program listed below generates the figures presented in Fig.~\ref{fig:wdc_results},
up to small fluctuations that are due to the use of pseudo-random numbers.
The program has been tested with Python version 2.7.2 (Windows 7) and Python 2.7.6 (Linux 3.13.0-24), with matplotlib installed.
Depending on the CPU used, it may take considerable time (an hour or more) to produce all 8 figures.
To check if the program runs it may be convenient to change
{\bf num\_data\_points = 50} and
{\bf num\_iter = 10000} into
{\bf num\_data\_points = 10} and
{\bf num\_iter = 1000}, respectively.
Progress of the calculation is indicated on the screen.

\begin{scriptsize}
\begin{lstlisting}[language=Python]
from math import *
from numpy import *
from random import random
import matplotlib.pyplot as plt
import matplotlib
import sys

delta_0 = pi*4/32
delta_1 = -pi*7/40
num_data_points = 50
num_iter = 10000

def heaviside(x):
 return 0.5 * (sign(x) + 1)

def inner_prod(x, y):
 return sum(x*conj(y))

def norm(z):
 return sqrt(inner_prod(z, z).real)

def half_wave_plate(v, theta):
 M_hwp = -1.0j*array([ [cos(2*theta), sin(2*theta)],
    [sin(2*theta), -cos(2*theta)]], dtype=complex)
 return dot(M_hwp, v)

class DLM:
 def __init__(self, num_input_ports = 2, num_output_ports = 1, gamma = 0.99):
  self.__gamma = gamma
  self.num_input_ports = num_input_ports

  if num_input_ports != 2:
   raise Exception("Number of input ports not equal to two not implemented!")

  self.x = zeros(num_input_ports)
  self.x[0] = random()
  self.x[1] = 1 - self.x[0]

  self.Y = zeros(num_input_ports*2, dtype=complex)
  for i in range(0, num_input_ports): # Initialize internal (unit) vectors
   r=random()
   self.Y[2*i] = r
   self.Y[2*i+1] = sqrt(1 - r**2)

 def learn(self, input_port):
  v = zeros(2)
  v[input_port] = 1
  self.x = self.__gamma * self.x + (1-self.__gamma)*v

class beam_splitter(DLM):
 def __init__(self, num_output_ports = 2, complex_field = True):
  DLM.__init__(self, 2, 2)
  self.num_output_ports = num_output_ports
  if complex_field == True:
   # Transformation matrices:
   self.T_0 = array([ [1, 0, 0, 0],
      [0, 0, 1, 0],
      [0, 1, 0, 0],
      [0, 0, 0, 1]])
   self.T_1 = 1/sqrt(2)*array([ [1, 1j, 0, 0],
       [1j, 1, 0, 0],
       [0, 0, 1, 1j],
       [0, 0, 1j, 1]])
   self.T = dot(self.T_0,dot(self.T_1, self.T_0))

  else:
   self.T = 1/sqrt(2)*array([ [1, 0, 0, -1],
       [0, 1, 1, 0 ],
       [0, -1, 1, 0],
       [1, 0, 0, 1]])

 def event(self, E_in, input_port):
  self.Y[input_port*2:input_port*2+2] = E_in

  self.learn(input_port)

  M = diag([sqrt(self.x[0]), sqrt(self.x[0]), sqrt(self.x[1]), sqrt(self.x[1])])
  g = dot(M, self.Y)
  g = dot(self.T, g)

  output_port = heaviside(random() - norm(g[0:2])**2)

  E_out = g[output_port*2:output_port*2+2]
  if norm(E_out) != 0:
   E_out = E_out / norm(E_out)

  return (E_out, output_port)

class polarizing_beam_splitter(beam_splitter):
 def __init__(self, num_output_ports = 2):
  beam_splitter.__init__(self, num_output_ports)
  self.T_1 = array([ [1, 0, 0, 0],
     [0, 1, 0, 0],
     [0, 0, 0, 1j],
     [0, 0, 1j, 0]])
  self.T = dot(self.T_0,dot(self.T_1, self.T_0))

class polarizer(beam_splitter):
 def __init__(self, num_output_ports = 2):
  beam_splitter.__init__(self, num_output_ports)
  self.__port = 0 # which port to accept messages
  self.R = 1.0/sqrt(2)*array([ [1, 1, 0, 0],   # Change of basis
      [1, -1, 0, 0],
      [0, 0, 1, 1],
      [0, 0, 1, -1]], dtype=complex)
  # Transformation matrices:
  self.T_1 = array([ [1, 0, 0, 0],
     [0, 1, 0, 0],
     [0, 0, 0, 0],
     [0, 0, 0, 0]])
  self.T = dot(self.R, dot(dot(self.T_0,dot(self.T_1, self.T_0)), linalg.inv(self.R)))

 def event(self, E_in, input_port):
  self.Y[input_port*2:input_port*2+2] = E_in

  self.learn(input_port)

  M = diag([sqrt(self.x[0]), sqrt(self.x[0]), sqrt(self.x[1]), sqrt(self.x[1])])
  g = dot(M, self.Y)
  g = dot(self.T, g)

  output_port = heaviside(random() - norm(g[0:2])**2)

  if output_port == self.__port:
   E_out = g[output_port*2:output_port*2+2]
   E_out = E_out / norm(E_out)
  else:
   E_out = zeros(2, dtype=complex)

  return (E_out, output_port)

for n in range(0,8):
 alpha = n*pi/16
 phi = zeros(num_data_points)

 def simulate(pol = True):
  bs1 = polarizing_beam_splitter()
  bs2 = polarizing_beam_splitter()
  bs2_2 = polarizing_beam_splitter()
  bs3 = polarizing_beam_splitter()
  bs3_2 = polarizing_beam_splitter()
  bs4 = polarizing_beam_splitter()
  bs4_2 = polarizing_beam_splitter()
  bs5 = polarizing_beam_splitter()
  bs5_2 = polarizing_beam_splitter()
  p = polarizer()

  N = zeros([num_data_points,2])
  sys.stdout.write("|")
  sys.stdout.flush()
  for i in range(0, num_data_points):
   sys.stdout.write(":")
   sys.stdout.flush()
   phi[i] = pi*i*8/num_data_points

   for k in range(0, num_iter):
    [E_out, detector_port] = bs1.event(array([cos(pi/4), sin(pi/4)], dtype=complex), 0)

    if detector_port == 1:
     E_out = E_out*exp(1j*phi[i])

    E_out = half_wave_plate(E_out, alpha)

    if detector_port == 1:
     E_out = half_wave_plate(E_out, pi/4)
     [E_out, layer_port] = bs2_2.event(E_out, 0)
    else:
     [E_out, layer_port] = bs2.event(E_out, 0)

    if layer_port == 0: # horizontal polarization
     if detector_port == 0:
      E_out = half_wave_plate(E_out, pi/4)


     [E_interm, port] = bs3.event(E_out, detector_port)
     E_interm = half_wave_plate(E_interm, pi/8)

     [E_out, detector_port] = bs4.event(E_interm, 0)

     if detector_port == 0:
      E_out = half_wave_plate(E_out, pi/4)

    else: # vertical polarization
     if detector_port == 1:
      E_out = half_wave_plate(E_out, pi/4)

     [E_interm, port] = bs3_2.event(E_out, detector_port)
     E_interm = half_wave_plate(E_interm, 0)
     [E_out, detector_port] = bs4_2.event(E_interm, 0)

     if detector_port == 1:
      E_out = half_wave_plate(E_out, pi/4)

    if detector_port == 0 and layer_port == 0:
     E_out[1] = E_out[1] * exp(1j*delta_1)
    elif detector_port == 1 and layer_port == 0:
     E_out[1] = E_out[1] * exp(1j*delta_0)

    if detector_port == 0:
     [E_out, output_port] = bs5.event(E_out, layer_port)
    else:
     [E_out, output_port] = bs5_2.event(E_out, layer_port)
    if pol:
     [E_out, output_port] = p.event(E_out, 0)
     if output_port == 0:
      N[i,detector_port] += 1
    else:
     N[i,detector_port] += 1

  sys.stdout.write("|\n")
  return N

 N = simulate()
 N_2 = simulate(False)

 # Theoretical formula:
 P_0c = 0.5 + 0.5*cos(2*alpha)**2*cos(phi)
 P_1c = 0.5 - 0.5*cos(2*alpha)**2*cos(phi)
 P_0plus = (cos(2*alpha)**2 * cos(phi/2)**2 + sin(2*alpha)**2/2.0 + \
            sqrt(2)*sin(2*alpha)*cos(2*alpha)*cos(phi/2)*cos(phi/2 + delta_0))/2.0
 P_1plus = (cos(2*alpha)**2 * sin(phi/2)**2 + sin(2*alpha)**2/2.0 - \
            sqrt(2)*sin(2*alpha)*cos(2*alpha)*sin(phi/2)*sin(phi/2 - delta_1))/2.0

 plt.clf()
 plt.plot(phi/(pi*2.0), double(N_2[:,0])/(N_2[:,0] + N_2[:,1]), \
          'rs', markeredgecolor='r', markersize = 9)
 plt.plot(phi/(pi*2.0), P_1c, 'r-', linewidth=1.5)
 plt.plot(phi/(pi*2.0), double(N[:,0])/(N[:,0]+N[:,1]), \
          'bo', markeredgecolor='b', markersize = 9)
 plt.plot(phi/(pi*2.0), P_1plus/(P_0plus+P_1plus), 'b--', linewidth=1.5)
 plt.xlabel(r"$\varphi/(2\pi)$")
 plt.ylabel("Event fraction")
 plt.axis([0.0,4.0, 0.0, 1.0])
 xticks = plt.gca().xaxis.get_major_ticks()
 xticks[0].label1.set_visible(False)

 matplotlib.rcParams.update({'font.size': 20})
 plt.tight_layout()
 plt.savefig("wdc" + str(n) + ".eps")
 print "Saved wdc" + str(n) + ".eps"

\end{lstlisting}
\end{scriptsize}

\section*{References}
\bibliographystyle{elsarticle-num}
\bibliography{../../../all13,qdc}
\end{document}